\documentclass{article}
\usepackage{spconf,amsmath,graphicx,url}
\usepackage{multirow}
\usepackage{array}
\usepackage{subcaption}

\title{A Framework for Robust Speaker Verification in Highly Noisy Environments Leveraging Both Noisy and Enhanced Audio}
\name{Adam Katav \quad Yair Moshe \quad Israel Cohen}
\address{Signal and Image Processing Laboratory (SIPL)\\
  Andrew and Erna Viterbi Faculty of Electrical \& Computer Engineering\\
  Technion -- Israel Institute of Technology, Haifa, Israel\\
  \url{https://sipl.ece.technion.ac.il}}

\begin{document}
%
\maketitle
\begin{abstract}

Recent advancements in speaker verification techniques show promise, but their performance often deteriorates significantly in challenging acoustic environments. Although speech enhancement methods can improve perceived audio quality, they may unintentionally distort speaker-specific information, which can affect verification accuracy. This problem has become more noticeable with the increasing use of generative deep neural networks (DNNs) for speech enhancement. While these networks can produce intelligible speech even in conditions of very low signal-to-noise ratio (SNR), they may also severely alter distinctive speaker characteristics.
To tackle this issue, we propose a novel neural network framework that effectively combines speaker embeddings extracted from both noisy and enhanced speech using a Siamese architecture. This architecture allows us to leverage complementary information from both sources, enhancing the robustness of speaker verification under severe noise conditions. Our framework is lightweight and agnostic to specific speaker verification and speech enhancement techniques, enabling the use of a wide range of state-of-the-art solutions without modification. Experimental results demonstrate the superior performance of our proposed framework.

\end{abstract}
\section{Introduction}
\begin{figure*}[t] 
    \centering
    \begin{subfigure}{0.32\textwidth} 
        \centering
        \includegraphics[width=\linewidth]{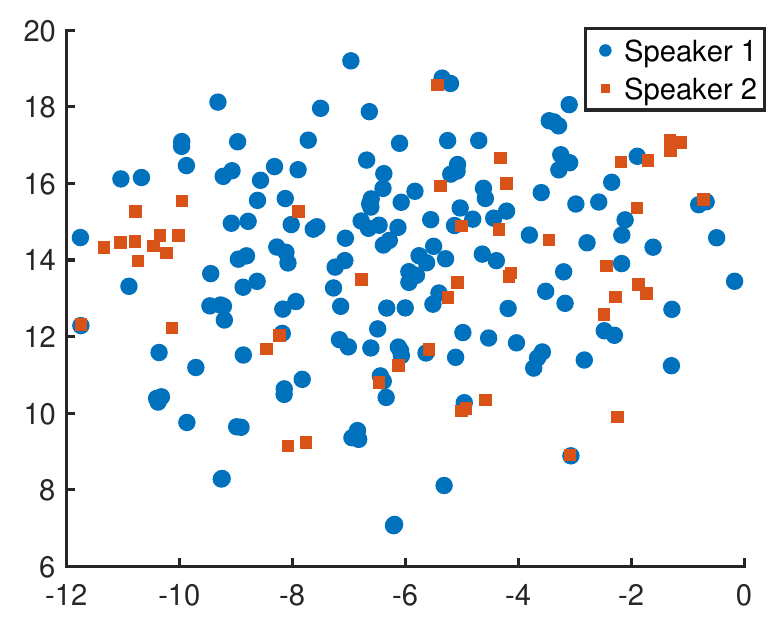}
        \caption{Noisy}
    \end{subfigure}
    \hfill
    \begin{subfigure}{0.32\textwidth}
        \centering
        \includegraphics[width=\linewidth]{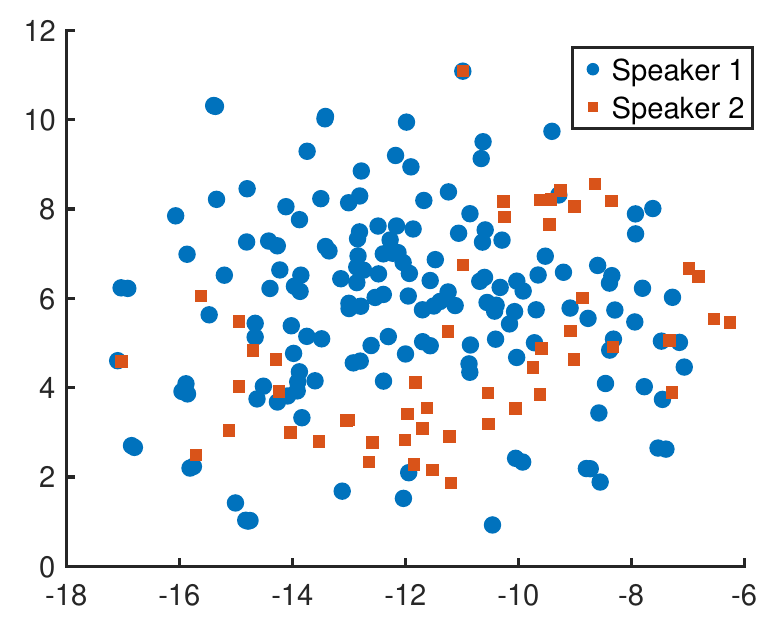}
        \caption{Enhanced}
    \end{subfigure}
    \hfill
    \begin{subfigure}{0.32\textwidth}
        \centering
        \includegraphics[width=\linewidth]{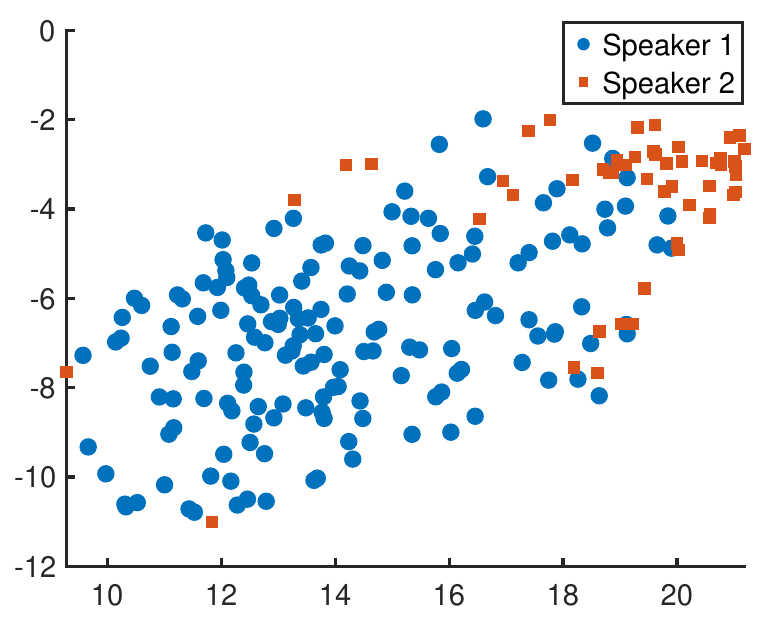}
        \caption{Proposed}
    \end{subfigure}
    \caption{t-SNE visualization of SpeakerNet~\cite{koluguri2020speakernet} embeddings extracted from two speakers in the VoxCeleb1~\cite{nagrani2020voxceleb} dataset with babble noise at an SNR of -15 dB: (a) without any processing, (b) enhanced using DeepFilterNet3~\cite{schroter2023deepfilternet}, and (c) generated using the proposed method, which combines embeddings from (a) and (b).}
    \label{fig:tsne}
\end{figure*}

Speaker verification aims to determine whether two audio samples originate from the same speaker. Typical speaker verification applications include voice authentication for personal smart devices, authenticating callers in call centers, securing access in telephone banking, and law enforcement investigations. Speaker verification systems rely on speaker embeddings, which are high-dimensional features representing a speaker’s vocal tract characteristics and speaking style. Once created, these embeddings can be compared to determine if two audio samples come from the same person \cite{bai2021speaker,jakubec2024deep}.

Over the last decade, the superior feature extraction capabilities of deep neural networks (DNNs) have made them the leading approach for speaker embedding. Among the pioneering works in this area was the introduction of the Time Delay Neural Network (TDNN) \cite{lang1990time}. One of the most successful architectures for speaker verification using this approach was developed in \cite{snyder2018x}. In this design, input mel-frequency coefficients (MFCC) are fed into a TDNN, and the resulting embeddings are termed x-vectors. SpeakerNet \cite{koluguri2020speakernet} is a lightweight model that uses an x-vector-based statistics pooling layer. It comprises residual blocks with 1-dimensional depth-wise separable convolutions, batch normalization, and ReLU layers. Another variant of the x-vector architecture with improved results is the emphasized channel attention, propagation, and aggregation TDNN (ECAPA-TDNN) \cite{desplanques2020ecapa}. This model integrates insights from computer vision, introducing enhancements such as 1D Res2Net modules with skip connections to improve the capture of temporal relationships, Squeeze-and-Excitation layers to emphasize informative channels, refining feature discrimination, and channel attention propagation and aggregation to further distribute attention weights across TDNN layers.

Speech recordings in the real world are often affected by background noise and reverberation. These disturbances make it challenging for a person to understand the speech and can also degrade the quality of extracted speaker embeddings, leading to inaccuracies in speaker verification. Speech enhancement plays an important role in speech signal processing as it aims to improve the intelligibility and quality of speech by removing noise from corrupted signals \cite{yuliani2021speech}. While it might seem intuitive that applying speech enhancement to noisy recordings would improve speaker embeddings, this is often untrue. The primary objective of speech enhancement is noise suppression, and it does not explicitly guarantee improvements in downstream tasks such as speaker verification. The artifacts and distortions introduced during the enhancement process can sometimes degrade speaker verification performance even further \cite{sadjadi2010assessment}. This issue has become more prominent recently as generative DNNs are increasingly used for speech enhancement. These models can produce superior speech quality and effectively enhance speech contaminated by higher noise levels. However, their generative nature can lead to significant distortions of the speaker’s intrinsic characteristics in the speech signal, especially under challenging noise conditions.

Several previous studies have investigated speaker verification in noisy conditions. Some of these works propose training a speech enhancement module specifically for speaker verification, using a tailored loss function \cite{kataria2020feature} or learning a mapping from noisy to clean speech embeddings \cite{plchot2016audio,cai2020within}. Other approaches focus on learning to extract robust speaker embeddings by separately capturing noise and speaker characteristics \cite{yu2021cam,sun2023noise}. A common strategy is to use a cascaded architecture, where instead of separately processing speech enhancement and speaker verification, the two modules are integrated into a single framework through joint optimization \cite{shon2019voiceid,kim2022extended,wu2023fused}. While existing works demonstrate enhanced noise robustness for speaker verification, they rely on training a dedicated speech enhancement module, a speaker verification module, or both. This is a major drawback as training state-of-the-art enhancement and verification modules often demands significant computational resources and access to large high-quality datasets, which may not be readily available. Our proposed framework offers a more practical solution by utilizing any such pre-trained enhancement or verification module out-of-the-box. This approach not only simplifies the design but also significantly reduces computation complexity. Unlike \cite{lee2023lc4sv}, which employs a learning-based interpolation agent to automatically determine the optimal linear combination of noisy and enhanced signals, our framework combines the speaker embeddings extracted from both sources. This nonlinear combination of embeddings is performed in a highly informative latent space, enabling improved performance without the added complexity of training an interpolation agent. Furthermore, our utilization of state-of-the-art speech enhancement techniques allows us to deliver reliable speaker verification performance even in severe noisy conditions where previous speaker verification methods fail and where speech enhancement methods introduce significant distortions to speaker characteristics.

\section{Robust Speaker Verification}
\begin{figure*}
\centering
\includegraphics[width=0.7\textwidth, trim=0 30 0 0, clip]{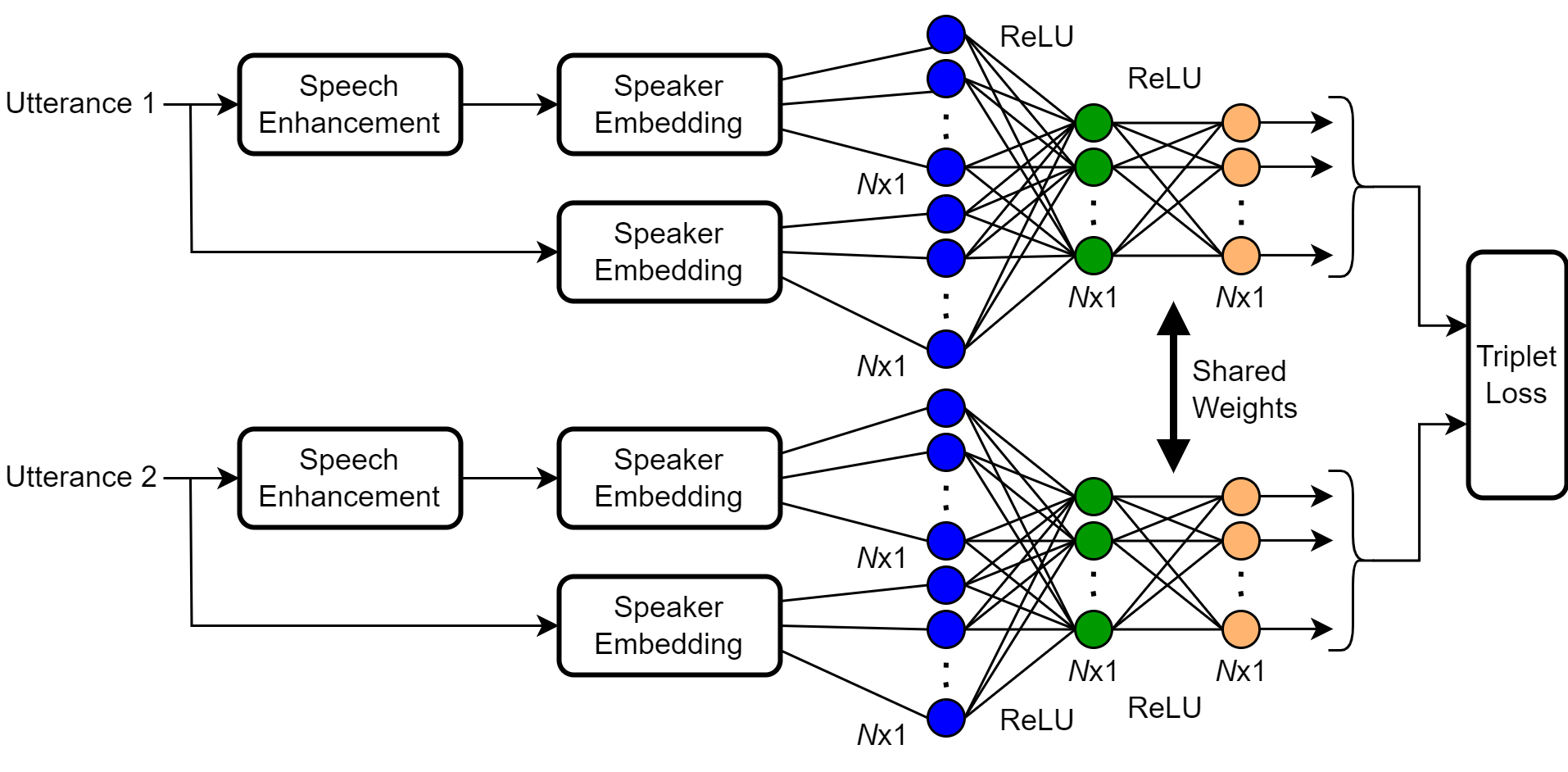}
\caption{Proposed Siamese architecture for learning robust speaker embeddings.}
\label{fig:block_diagram}
\end{figure*}


The solution we propose is based on the understanding that speaker embeddings extracted from noisy speech signals and their corresponding enhanced signal embeddings provide complementary information. By combining this complementary information, we can create a more robust embedding that is resistant to noise. In conditions with relatively low noise, the noisy embedding is expected to be more informative because it is less likely to be affected by significant noise artifacts. On the other hand, in challenging noise conditions, the enhanced embedding is likely to provide more valuable information, benefiting from the noise reduction achieved through the enhancement process. We suggest training a neural network to effectively combine these two types of embeddings, taking into account various types and levels of noise. This network can adaptively adjust the contributions of each embedding based on the noise level perceived in the input speech. As a result, this approach can enhance speaker verification performance across different noise scenarios. Figure~\ref{fig:tsne} illustrates this concept with a t-SNE visualization of embeddings from noisy utterances of two speakers in the VoxCeleb1 dataset. The figure shows that, compared to our approach where noisy and enhanced embeddings of each utterance are combined, the noisy and enhanced embeddings of the two speakers are more distinctly separable. This facilitates more accurate speaker verification.

As illustrated in Fig.~\ref{fig:block_diagram}, our framework is based on Siamese neural networks. These neural networks are specifically designed to compare pairs of inputs and assess their degree of similarity. This capability has proven invaluable in face recognition, image matching, and fingerprint verification applications. A Siamese architecture involves feeding two inputs into the network: a query input and a reference input. Each input is processed by identical subnetworks with the same weights and parameters, resulting in extracted features. A distance function is then applied to calculate the distance between the feature vectors of the two inputs. The computed distance is a metric for determining the similarity between the inputs, with a smaller distance indicating a higher degree of similarity. In our case, the inputs to the network are two speech utterances. Two embeddings of size $N\times1$ are extracted for each speech utterance: one from the noisy utterance and one from the enhanced utterance. The two embeddings of each utterance are combined using a 3-layer MLP (Multi-Layer Perceptron). The first layer is of size $2N\times1$, while the other two are each of size $N\times1$. Between the layers, a ReLU activation function is applied. The MLP learns nonlinear relationships between the two input embeddings and performs dimensionality reduction to output a robust speaker embedding of size $N\times1$. The two MLPs, which follow the Siamese architecture, share the same structure and weights. These weights are optimized based on the similarity between the two robust embeddings, as determined by the loss function. Note that the 3-layer MLP is significantly more straightforward to train than deeper networks and requires less computational resources. 

This framework offers the flexibility to employ any speech enhancement and speaker embedding method, enabling the seamless integration of state-of-the-art techniques out-of-the-box. By leveraging cutting-edge speech enhancement methods, we can deliver reliable speaker verification performance even in highly challenging noisy environments, where other speaker verification methods struggle and speech enhancement methods introduce significant distortions to speaker characteristics.

We employ a triplet loss function~\cite{schroff2015facenet} commonly used for recognition and verification tasks. This loss aims to learn a distance metric that effectively distinguishes between similar and dissimilar examples. We use a variant of triplet loss based on cosine distance to account for the magnitude-invariance property often desired when comparing speaker embeddings. This loss is defined as:
\begin{equation}
    \mathcal{L}(A,P,N)=\max(0,d(A,P)-d(A,N)+\alpha)
\end{equation}
where $A$ is an anchor utterance, $P$ is a utterance from the same speaker as $A$, and $N$ is a utterance from a different speaker. The margin $\alpha$ controls the separation between positive and negative pairs. The function $d(X,Y)$ is the cosine distance between $X$ and $Y$, derived from the cosine similarity that measures the angle $\theta$ between the two vectors:
\begin{equation}
    d(X,Y) = 1-\cos(\theta) = 1 - \frac{X \cdot Y}{||X||||Y||}\,.
\end{equation}
Triplet loss minimizes the distance between an anchor and its positive examples (utterances from the same speaker), while maximizing the distance between the anchor and its negative examples (utterances from different speakers). This approach encourages the model to place utterances from the same speaker closer together in the embedding space, while ensuring that utterances from different speakers are positioned further apart.

\begin{table*}[tb]
\caption{Speaker verification results using our proposed method, compared to verification on noisy signals and noisy signals enhanced with DeepFilterNet3~\cite{schroter2023deepfilternet}. The evaluation was performed using SpeakerNet~\cite{koluguri2020speakernet} and ECAPA-TDNN~\cite{desplanques2020ecapa} speaker embeddings, with noise from the MUSAN corpus~\cite{snyder2015musan}, covering three noise types at varying SNR levels. Performance is measured using the Equal Error Rate (EER), with the best results highlighted in bold.}
\label{tab:results}
\centering
\begin{tabular}{|c|c||>{\centering\arraybackslash}p{1cm}|>{\centering\arraybackslash}p{1cm}|>{\centering\arraybackslash}p{1cm}||>{\centering\arraybackslash}p{1cm}|>{\centering\arraybackslash}p{1cm}|>{\centering\arraybackslash}p{1cm}|}
\hline
\multicolumn{2}{|c||}{} & \multicolumn{3}{c||}{SpeakerNet} & \multicolumn{3}{c|}{ECAPA-TDNN} \\
\hline
Type & SNR & Noisy & Enhc & Ours & Noisy & Enhc & Ours \\ \hline\hline

\multirow{5}{*}{\raggedright Noise} & 0 & \textbf{9.70} & 13.45 & 13.17 & \textbf{3.31} & 7.68 & 9.43 \\ \cline{2-8}
& -5 & 16.39 & 18.18 & \textbf{15.67} & \textbf{5.50} & 12.22 & 10.76 \\ \cline{2-8}
& -10 & 26.19 & 25.12 & \textbf{19.31} & \textbf{12.53} & 19.08 & 14.74 \\ \cline{2-8}
& -15 & 34.71 & 32.77 & \textbf{25.21} & 23.23 & 26.91 & \textbf{21.48} \\ \cline{2-8}
& -20 & 41.66 & 39.69 & \textbf{33.00} & 31.86 & 34.27 & \textbf{29.90} \\ \hline\hline

\multirow{5}{*}{\raggedright Music} & 0 & \textbf{12.42} & 14.62 & 15.33 & \textbf{4.96} & 8.31 & 12.19 \\ \cline{2-8}
& -5 & 23.73 & 22.80 & \textbf{19.48} & \textbf{13.61} & 16.28 & 16.37 \\ \cline{2-8}
& -10 & 36.82 & 33.72 & \textbf{27.37} & 28.41 & 27.38 & \textbf{24.49} \\ \cline{2-8}
& -15 & 44.46 & 41.86 & \textbf{36.37} & 41.11 & 38.09 & \textbf{34.96} \\ \cline{2-8}
& -20 & 48.69 & 47.39 & \textbf{44.05} & 47.90 & 45.77 & \textbf{43.88} \\ \hline\hline

\multirow{5}{*}{\raggedright Babble} & 0 & \textbf{20.24} & 24.79 & 21.89 & \textbf{19.13} & 23.16 & 24.04 \\ \cline{2-8}
& -5 & 32.64 & 36.59 & \textbf{26.82} & 34.89 & 37.63 & \textbf{31.43} \\ \cline{2-8}
& -10 & 43.85 & 44.77 & \textbf{32.30} & 45.15 & 45.50 & \textbf{38.47} \\ \cline{2-8}
& -15 & 46.72 & 47.48 & \textbf{37.21} & 48.27 & 47.78 & \textbf{42.18} \\ \cline{2-8}
& -20 & 48.31 & 48.38 & \textbf{41.73} & 48.77 & 48.62 & \textbf{45.38} \\ \hline
\end{tabular}
\end{table*}

\section{Results}
Our proposed framework was trained and evaluated using the VoxCeleb1 dataset \cite{nagrani2020voxceleb}, a collection of celebrity utterances extracted from YouTube videos. The VoxCeleb1 training set comprises 148,642 utterances from 1,211 speakers, while the test set contains 4,874 utterances from 40 speakers. To simulate real-world noise conditions, we augmented both training and test sets with recordings from the MUSAN corpus \cite{snyder2015musan}, which includes three categories of noises: 6 hours of general noise, such as DTMF tones, thunder, footsteps, paper rustling, and animal noises, 42 hours of music, and 60 hours of speech babble noise. We divided the MUSAN corpus into two disjoint sets: one for training set augmentation and the other for test set augmentation. Each utterance in the training set was corrupted with a randomly selected signal-to-noise ratio (SNR) between 0 and -20~dB. The test set was evaluated at SNR levels of \{0, -5, -10, -15, -20\} dB. These SNR levels are lower than those typically found in previous literature. Our framework is especially beneficial at these lower SNR levels due to its innovative use of state-of-the-art speech enhancement and speaker verification models out-of-the-box. Our model was trained using the AdamW optimizer with a batch size of 32, a learning rate of $10^{-3}$, and a triplet margin parameter $\alpha=0.25$. Since it is lightweight, it was trained for just 10 minutes on an NVIDIA RTX 3090.

Table~\ref{tab:results} summarizes the results of our proposed method, compared to speaker verification using noisy and enhanced embeddings, where the enhancement was performed with DeepFilterNet3~\cite{schroter2023deepfilternet}. The evaluation was conducted using speaker embeddings from SpeakerNet~\cite{koluguri2020speakernet} and ECAPA-TDNN~\cite{desplanques2020ecapa}. Notably, ECAPA-TDNN demonstrates greater robustness to noise compared to SpeakerNet, typically resulting in better performance at lower SNR levels. At SNR = 0, the noisy embedding achieves the best verification performance. However, at lower SNRs, the enhanced embedding occasionally outperforms it, while our proposed method consistently delivers the best results in these lower SNRs.

\section{Conclusions}

This paper presented a novel neural network framework for robust speaker verification in challenging acoustic environments. The proposed Siamese architecture effectively integrates speaker embeddings from both noisy and enhanced speech, leveraging their complementary information to improve verification performance. By utilizing state-of-the-art, pre-trained speech enhancement and speaker verification models, the framework eliminates the need for task-specific training, making it both practical and computationally efficient.
Experimental results demonstrate the superiority of the proposed approach across diverse noise conditions, particularly in highly degraded environments where conventional speaker verification methods often fail. Moreover, the framework is inherently flexible and agnostic to specific speech enhancement and speaker verification techniques, enabling seamless integration with future advancements in these fields. 

\section{Acknowledgment}
The authors thank Ram Binshtock, Sahar Zeltzer, and David Portal for their contributions in the early stages of this work as part of the Signal Processing Cup 2024 Challenge.
This research was supported by the Israel Science Foundation (grant no. 1449/23) and the Pazy Research Foundation. 

\bibliographystyle{IEEEbib}
\bibliography{references}

\end{document}